\documentclass[aps,prd,preprint,superscriptaddress,amsmath,amssymb,showpacs]{revtex4-1}
\usepackage{dcolumn}
\usepackage{graphicx}
\usepackage{float}
\usepackage{physics}
\usepackage[colorlinks=true,allcolors=blue]{hyperref}
\usepackage{mathrsfs}
\usepackage{inputenc}
\begin{document}

\title{Holographic dissociation of tetraquark and hybrid meson in the spinning plasma}

\author{Zhenqiang Ren }
\email{renzhq@hyit.edu.cn}
\affiliation{Faculty of Electronic Information Engineering, Huai'an University, Huai'an 223003, China}

\author{Xiang Gao }
\email{gaoxiang@zknu.edu.cn}
\affiliation{School of Physics and Telecommunications Engineering, Zhoukou Normal University, Zhoukou 466001, China}

\author{Zhou-Run Zhu }
\email{zhuzhourun@zknu.edu.cn}
\affiliation{School of Physics and Telecommunications Engineering, Zhoukou Normal University, Zhoukou 466001, China}

\begin{abstract}
We explore the dissociation of the tetraquark \(Z_c\) and the hybrid meson \(\pi_1\) in a spinning quark-gluon plasma described by the planar limit of a spinning Myers-Perry black hole. Both states are treated within one holographic framework, using a non-quadratic dilaton profile whose parameters are fixed by the \(Z_c\) and \(\pi_1\) mass spectra. Spectral functions are computed in the longitudinal and transverse channels. The boost parameter and temperature both lower the peak height and broaden the width, and the effect is stronger in the transverse channel. In both channels, the heavier 2S states melt before the 1S states. Furthermore, the two exotic channels display distinct dissociation behaviors within the same effective framework. This provides a unified holographic approach to compare how different exotic states behavior, without requiring a dynamical calculation of their binding energies.
\end{abstract}

\maketitle

\section{Introduction}\label{sec:01_intro}

Heavy-ion collisions provide access to the quark-gluon plasma (QGP) \cite{Arsene:2004fa,Adcox:2004mh,Back:2004je,Adams:2005dq}. Heavy quarkonium, a bound state of a heavy quark and its antiquark, is among the most widely used probes of this strongly interacting medium. The color interaction between the quark and antiquark is screened by the surrounding medium, and this screening can weaken the binding enough to dissolve the state. This mechanism suppresses quarkonium production, and the suppression has become a standard signature of deconfinement \cite{Matsui:1986dk}. The heavy quarkonium spectra reflects the properties of the QGP. The spectral function resonance peaks give the in-medium mass and width, and the broadening or disappearance of a peak signals that the bound state has melted.

The gauge/gravity duality \cite{Maldacena:1997re,Witten:1998qj,Gubser:1998bc} provides a non-perturbative method to strongly coupled gauge theories, mapping a gravitational theory in asymptotically AdS space. Heavy flavor physics has been explored from holography. Early holographic work addressed the thermal behavior of scalar glueballs and scalar mesons \cite{Colangelo:2009ra,Miranda:2009uw}, and subsequent studies examined magnetic field effects \cite{Dudal:2014jfa,Braga:2018zlu,Zhao:2021ogc}, finite temperature and chemical potential \cite{Mamani:2013ssa,Braga:2016wkm,Braga:2017oqw,Braga:2017bml,Mamani:2022qnf}, masses and decay constants \cite{Braga:2015jca,Braga:2015lck}, quasinormal modes \cite{Braga:2019yeh,Braga:2019xwl,Zhao:2023yry}, rotating plasmas \cite{Zhao:2023pne,Braga:2023fac,Zhu:2024uwu}, and anisotropic backgrounds \cite{Chang:2024ksq}. Spectral functions in Gauss-Bonnet gravity can be found in \cite{Zhu:2024ddp}.

Exotic states have drawn growing interest because their internal structure is not captured by a simple quark-antiquark pair. The tetraquark \(Z_c(3900)^+\), observed by BESIII in \(e^+e^-\to\pi^+\pi^- J/\psi\) \cite{BESIII:2015cld}, carries quantum numbers \(I^G(J^{PC})=1^+(1^{+-})\) and has a minimal valence content of \(c\bar{c}u\bar{d}\). Its interpretation as a tetraquark, a hadronic molecule, or a threshold effect is still debated \cite{Lebed:2016hpi,Guo:2017jvc}. The hybrid meson \(\pi_1(1400)\) has \(I^G(J^{PC})=1^-(1^{-+})\), quantum numbers forbidden in the constituent quark model; its interpolating field must contain a valence gluon \cite{Meyer:2015eta}. These states are not merely spectroscopic curiosities. Their binding involves multiquark correlations or gluonic flux, and their response to a thermal medium may therefore differ from that of ordinary quarkonia. A holographic calculation found that the \(\pi_1\) hybrid dissociates at a lower temperature than light \(q\bar{q}\) mesons \cite{Bellantuono:2014lra}. More recently, the non-quadratic dilaton model of \cite{MartinContreras:2020cyg,MartinContreras:2021bis} was combined with a self-consistent Einstein-Maxwell-dilaton framework to reproduce the mass spectra of \(Z_c\) and \(\pi_1\) \cite{Toniato:2025gts}, giving a tested set of holographic parameters for these two channels. In this approach the structural difference between the two states is encoded in the fitted parameters, rather than in a dynamical calculation of their binding. What has not been examined is how these states behave in a spinning medium.

Non-central heavy-ion collisions carry a nonzero angular momentum, part of which can be transferred to the QGP \cite{Liang:2004ph,Becattini:2007sr,Baznat:2013zx,STAR:2017ckg,Jiang:2016woz}. Rotation may affect the confinement-deconfinement transition \cite{Chen:2020ath,Braga:2022yfe,Zhao:2022uxc}, parton energy loss \cite{Hou:2021own,Chen:2022obe,Chen:2023yug}, and the entropy of heavy quarkonium \cite{Wu:2022ufk}. Other significative researches are detailed in \cite{McInnes:2014haa,McInnes:2016dwk,Cai:2023cjl,Zhang:2023psy,Zhao:2024ipr,Wang:2026mwm,Zhao:2026yxk}. In \cite{Zhao:2023pne,Braga:2023fac}, a rotating background was built from a local Lorentz transformation \cite{Erices:2017izj,BravoGaete:2017dso,Awad:2002cz}, but that construction covers only a small neighbourhood around a rotating radius and a domain narrower than $2\pi$. The spinning Myers-Perry black hole \cite{Hawking:1998kw,Gibbons:2004ai,Gibbons:2004js,Garbiso:2020puw} provides a global rotating background. In the planar limit, the dual gauge theory lives on flat space $\mathbb{R}^{3,1}$, and stability requires $a<0.75L$ \cite{Garbiso:2020puw}. The heavy and exotic spectral functions were developed recently in a gluon-condensate background \cite{Zhao:2026bsc,Wang:2026jfn}.

In this paper, we compute the spectral functions of $Z_c$ and $\pi_1$ in the spinning Myers-Perry black hole background. It remains unclear how exotic states behavior in the spinning background. The results presented below show that the dissociation temperatures of the \(Z_c\) and \(\pi_1\) channels differ significantly from each other. In the planar limit the angular velocity $\Omega$ is related to the boost parameter $a$ as \(\Omega = a/L^2\). Longitudinal and transverse channels are treated separately. We find that both the boost parameter and the temperature suppress the peak height and broaden the peak width, enhancing the dissociation of $Z_c$ and $\pi_1$, with a stronger effect in the transverse orientation.

The paper is organized as follows. In Sec.~\ref{sec:02}, we briefly review the spinning Myers-Perry black hole. In Sec.~\ref{sec:03}, we study the spectral function of $Z_c$ and $\pi_1$ in spinning Myers-Perry black hole. The conclusion and discussion are given in Sec.~\ref{sec:04}.

\section{Spinning Myers-Perry black hole background}\label{sec:02}

We begin with a brief review of the spinning black hole background introduced by Ref. \cite{Hawking:1998kw}, which will be used for the tetraquark $Z_c$ and hybrid meson $\pi_1$ spectral functions. The metric takes the form
 \begin{equation}
\label{eqc1}
\begin{split}
ds^{2}=&-\frac{\Delta}{\rho^{2}}(dt_{H}-\frac{a\sin^{2}\theta_{H}}{\Xi_{a}}d\phi_{H}-\frac{b\cos^{2}\theta_{H}}{\Xi_{b}}d\psi_{H})^{2}+\frac{\Delta_{\theta_{H}}\sin^{2}\theta_{H}}{\rho^{2}}(adt_{H}-\frac{r_{H}^{2}+a^{2}}{\Xi_{a}}d\phi_{H})^{2}\\
 &+\frac{\Delta_{\theta_{H}}\cos^{2}\theta_{H}}{\rho^{2}}(bdt_{H}-\frac{r_{H}^{2}+b^{2}}{\Xi_{b}}d\psi_{H})^{2}+\frac{\rho^{2}}{\Delta}dr_{H}^{2}-\frac{\rho^{2}}{\Delta_{\theta_{H}}}d\theta_{H}^{2}\\
 &+\frac{1+\frac{r_{H}^{2}}{L^{2}}}{r_{H}^{2}\rho^{2}}(abdt_{H}-\frac{b\left(r^{2}+a^{2}\right)\sin^{2}\theta_{H}}{\Xi_{a}}d\phi_{H}-\frac{a\left(r^{2}+b^{2}\right)\cos^{2}\theta_{H}}{\Xi_{b}}d\psi_{H})^{2},
 \end{split}
\end{equation}
with
 \begin{equation}
\label{eqc11}
\begin{split}
\Delta=\frac{1}{r_{H}^{2}}(r_{H}^{2}+a^{2})(r_{H}^{2}+b^{2})(1+\frac{r_{H}^{2}}{L^{2}})-2M,\\
 \Delta_{\theta_{H}}=1-\frac{a^{2}}{L^{2}}\cos^{2}\theta_{H}-\frac{b^{2}}{L^{2}}\sin^{2}\theta_{H},\\
 \rho=r_{H}^{2}+a^{2}\cos^{2}\theta_{H}+b^{2}\sin^{2}\theta_{H},\\
 \Xi_{a}=1-\frac{a^{2}}{L^{2}},\\
\Xi_{b}=1-\frac{b^{2}}{L^{2}},\\
 \end{split}
\end{equation}
where \(t_H\) is the time, \(L\) the AdS radius, and \(r_H\) the AdS radial coordinate. The angular Hopf coordinates are \((\phi_H, \psi_H, \theta_H)\). The parameters \(a\) and \(b\) denote the angular momenta. In this paper, we consider the spinning Myers-Perry black hole with \(a=b\) \cite{Gibbons:2004ai,Gibbons:2004js}.

For convenience, one can introduce a new set of coordinates and reparameterize the mass \cite{Murata:2008xr}.
 \begin{equation}
\label{eqc111}
\begin{split}
t=t_{H},\\
 r^{2}=\frac{a^{2}+r_{H}^{2}}{1-\frac{a^{2}}{L^{2}}},\\
\theta=2\theta_{H},\\
 \phi=\phi_{H}-\psi_{H},\\
\psi=-\frac{2at_{H}}{L^{2}}+\phi_{H}+\psi_{H},\\
b=a,\\
\mu=\frac{M}{(L^{2}-a^{2})^{3}}.\\
 \end{split}
\end{equation}

The metric of Eq.~\eqref{eqc1} takes the form
 \begin{equation}
\label{eqc2}
\begin{split}
ds^{2}=&-(1+\frac{r^{2}}{L^{2}})dt^{2}+\frac{dt^{2}}{G(\text{r)}}+\frac{r^{2}}{4}((\sigma^{1})^{2}+(\sigma^{2})^{2}+(\sigma^{3})^{2})+\frac{2\mu}{r^{2}}(dt+\frac{a}{2}\sigma^{3})^{2},
 \end{split}
\end{equation}
with
 \begin{equation}
\label{eqc3}
\begin{split}
G(r)=1+\frac{r^{2}}{L^{2}}-\frac{2\mu(1-\frac{a^{2}}{L^{2}})}{r^{2}}+\frac{2\mu a^{2}}{r^{4}},\\
\mu=\frac{r_{h}^{4}(L^{2}+r_{h}^{2})}{2L^{2}r_{h}^{2}-2a^{2}(L^{2}+r_{h}^{2})},\\
\sigma^{1}=-sin\psi dtd\theta+cos\psi sin\theta d\phi,\\
\sigma^{2}=cos\psi d\theta+sin\psi sin\theta d\phi,\\
\sigma^{3}=d\psi+cos\theta d\phi,\\
 \end{split}
\end{equation}
where
\begin{equation}
\label{eqc4}
\begin{split}
&-\infty<t<\infty,\ r_{h}<r<\infty,\ 0\leq\theta\leq\pi,\\
&0\leq\phi\leq2\pi,\ 0\leq\psi\leq4\pi.
 \end{split}
\end{equation}

To obtain the planar black brane, one can apply the coordinate transformation \cite{Garbiso:2020puw}
 \begin{equation}
\label{eqc5}
\begin{split}
t=\tau,\\
\frac{L}{2}(\phi-\pi)=x_{1},\\
\frac{L}{2}tan(\theta-\frac{\pi}{2})=x_{2},\\
\frac{L}{2}(\psi-2\pi)=x_{3},\\
r=\tilde{r},\\
 \end{split}
\end{equation}
where \((\tau, \widetilde{r}, x, y, z)\) stand for the new coordinates, which are subsequently rescaled by powers of \(\beta\).
\begin{equation}
\label{eqc6}
\begin{split}
\tau\rightarrow\beta^{-1}\tau,\\
x_{1}\rightarrow\beta^{-1}x_{1},\\
x_{2}\rightarrow\beta^{-1}x_{2},\\
x_{3}\rightarrow\beta^{-1}x_{3},\\
\tilde{r}\rightarrow\beta\tilde{r},\\
\tilde{r_{h}}\rightarrow\beta\tilde{r_{h}}.(\beta\rightarrow \infty)\\
 \end{split}
\end{equation}

This yields the Schwarzschild black brane metric, boosted in the \(\tau\)-\(x_3\) plane \cite{Garbiso:2020puw}
\begin{equation}
\label{eqc7}
\begin{split}
ds^{2}=&\frac{r^{2}}{L^{2}}(-d\tau^{2}+dx_{1}^{2}+dx_{2}^{2}+dx_{3}^{2}+\frac{r_{h}^{4}}{r^{4}(1-\frac{a^{2}}{L^{2}})}(d\tau+\frac{a}{L}dx_{3})^{2})+\frac{L^{2}r^{2}}{r^{4}-r_{h}^{4}}dr^{2},
 \end{split}
\end{equation}
where the metric reduces to the Schwarzschild black brane when boost parameter \(a=0\).

To simplify the calculation, we replace the holographic coordinate \(r\) by \(z\), with \(r=1/z\).
\begin{equation}
\label{eqc8}
\begin{split}
ds^{2}=& \frac{1}{z^{2}L^{2}}[-d\tau^{2}+dx_{1}^{2}+dx_{2}^{2}+dx_{3}^{2}+\frac{z^{4}}{z_{h}^{4}(1-\frac{a^{2}}{L^{2}})}(d\tau+\frac{a}{L}dx_{3})^{2}]+\frac{L^{2}z_{h}^{4}}{z^{2}(z_{h}^{4}-z^{4})}dz^{2}.
 \end{split}
\end{equation}

The temperature is obtained as \cite{Garbiso:2020puw}
\begin{equation}
\label{eqb1}
T=\frac{\sqrt{L^{2}-a^{2}}}{z_{h}\pi L^{3}},
\end{equation}
where $z_h$ represents the horizon.

The metric of Eq.~\eqref{eqc8} represents the local planar limit of the globally rotating Myers-Perry geometry. In the planar limit, the original angular motion is perceived locally as a tangential plasma flow, with \(x_{3}\) labeling the local flow direction. In the planar limit the angular velocity is \(\Omega = a/L^2\) \cite{Garbiso:2020puw}. The fluid flow lines for an embedding in \(\mathbb{R}^3\) may be pictured as lying on a torus \cite{Amano:2023bhg}. The ultra-local behavior of the flow is that of a uniform motion, corresponding to uniformly rotating circles of a Hopf fibration of the 3-sphere that forms the spatial part of the four-dimensional spacetime \cite{Amano:2023bhg}. To keep the temperature positive, boost parameter \(a\) must satisfy \(a<L\). The momentum diffusion coefficient further shows that the spinning black brane is stable for \(a<0.75L\) \cite{Garbiso:2020puw}. We therefore take \(a<0.75L\) in the numerical calculations.

Ref. \cite{Amano:2023bhg} examines the hydrodynamics of the Myers-Perry black hole. The authors find that low-energy fluctuations treat the rotating fluid as a boost with parameter \(a\). The frequencies and momenta of collective excitations around the rotating state are small relative to the temperature and angular momentum scales. This lets us treat the rotating fluid as a boosted fluid on its own scale. The result suggests that at intermediate to high temperatures, the plasma is well approximated by boosted-fluid models. The hydrodynamic modes, Lyapunov exponents, and butterfly velocities all indicate that the flow is uniform. Hence, for an embedding in \(\mathbb{R}^3\), the fluid flow lies on a torus and is uniform in the Myers-Perry black hole. We therefore take the dissociation effect to be homogeneous in this particular spinning Myers-Perry black hole. Within this approximation the medium is homogeneous, so the results describe the local response of an exotic state to the tangential flow, not the full spatial dependence of a finite fireball.

\section{Spectral function in the spinning Myers-Perry black hole background}\label{sec:03}

Using the phenomenological model of Ref.~\cite{Braga:2018zlu}, we study the spectral functions of tetraquark and hybrid meson in the spinning Myers-Perry black hole background. The required formulas follow from Ref.~\cite{Braga:2018zlu}. For convenience, the metric (\ref{eqc8}) is recast in a general form:
\begin{equation}
\begin{split}
\label{eqa1}
ds^{2}=& -g_{\tau \tau} d\tau^2 + g_{xx_{1}} (dx_{1}^{2}+ dx_{2}^{2})+g_{xx_{3}}dx_{3}^{2}+g_{\tau x_3}d\tau dx_3 + g_{x_3 \tau}dx_3d\tau+ g_{zz} dz^{2}.
 \end{split}
\end{equation}

One can describe the exotic mesons by a vector field \(V_m=(V_\mu,V_z)\), dual to the gauge-theory current \(J^\mu=\bar{\Psi}\gamma^\mu\Psi\). Its bulk action reads \cite{Braga:2018zlu}
\begin{equation}\label{eq:x1}
I= -\int d^{4}x dz \frac{Q}{4}F_{mn}F^{mn},
\end{equation}
where $F_{mn}=\partial_m V_n -\partial_n V_m$ and $Q=\frac{\sqrt{-g}}{e^{\phi(z)}g^2_5}$. We set $g_5=1$ in the calculations.  The dilaton field $\phi(z)$ is used to characterise vector mesons\cite{MartinContreras:2021bis},
\begin{equation}\label{eq:x4}
\phi(z)= (\kappa z)^{2-\alpha} + Mz + \tanh\left(\frac{1}{Mz} -\frac{\kappa}{\sqrt{\Gamma}}\right),
\end{equation}
where \(\kappa\), \(M\), \(\Gamma\), and \(\alpha\) are fixed by the nonlinear Regge trajectory. Their values for the tetraquark \(Z_c\) and the hybrid meson \(\pi_1\) are listed below \cite{Toniato:2025gts}
\begin{equation}
\label{eqav5}
\begin{split}
&  \kappa_{Z_c}=1.75GeV, M_{Z_c}=1.44GeV, \sqrt{\Gamma_{Z_c}}=0.30GeV, \alpha_{Z_c} = 0.539;\\
 & \kappa_{\pi_1}=0.468GeV, M_{\pi_1}=0.20GeV, \sqrt{\Gamma_{\pi_1}}=0.12GeV, \alpha_{\pi_1} = 0.034.
 \end{split}
\end{equation}

Following Ref.~\cite{Toniato:2025gts}, both channels are treated within one effective scheme. For \(Z_c\), the mass scale is taken from a hadronic-molecule picture, \(\bar{m}_{\mathrm{tetra}} = 0.283\,m_{J/\psi} + 0.717\,m_{\rho}\). For \(\pi_1\), it is taken from a hybrid picture with valence quarks and a constituent gluon, \(\bar{m}_{\mathrm{hybrid}} = 0.497\,m_u + 0.497\,m_d + 0.006\,m_G\). In each case the averaged mass and the assigned scale dimension fix \((\kappa, M, \sqrt{\Gamma}, \alpha)\) in \(\phi(z)\). The resulting mass spectra reproduce the experimental values of \(Z_c(3900)\), \(Z_c(4200)\), \(\pi_1(1400)\), and \(\pi_1(1600)\) at the few-percent level, which provides a quantitative basis for the parametrization used here. This is an effective description rather than a dynamical calculation. The fitted mass scales should therefore be understood as effective parameters that reproduce the spectral features of the two channels within this background, not as microscopic binding energies of the physical states.

One can compute the spectral function using the membrane paradigm \cite{Iqbal:2008by}. The equation of motion follows from Eq.~(\ref{eq:x1}).
\begin{equation}
\label{eqa8}
 \partial^m (Q F_{mn})=0 ,
\end{equation}
and the conjugate momentum to the gauge field on constant-\(z\) slices reads
\begin{equation}
\label{eqa9}
 j^\mu= -Q F^{z\mu}.
\end{equation}

One can take the plane wave to propagate along the \(x_3\) direction. The equations of motion (\ref{eqa8}) then split into longitudinal and transverse channels. Longitudinal fluctuations lie in the \((\tau,x_3)\) plane, while transverse fluctuations lie in the \((x_1,x_2)\) plane. Thus the longitudinal components of Eq.~\eqref{eqa8} read
\begin{equation}
\label{eqav8}
\begin{split}
&-\partial_z j^\tau-Q(g^{xx_3}g^{\tau \tau}+g^{x_3 \tau}g^{\tau x_3 })\partial_{x_3}F_{x_3 \tau}=0,\\
& -\partial_z j^{x_3}+ Q(g^{\tau \tau}g^{xx_3}+g^{\tau x_3 }g^{x_3 \tau})\partial_{\tau}F_{x_3 \tau}=0,\\
& \partial_{x_3} j^{x_3}+\partial_\tau j^\tau =0.
\end{split}
\end{equation}

The Bianchi identity then gives
\begin{equation}
\label{eqa91}
\partial_z F_{x_3 \tau}-\frac{g_{zz}}{Q}\partial_\tau[g_{xx_3}j^{x_3}+g_{x_3 \tau}j^{\tau}]-\frac{g_{zz}}{Q}\partial_{x_3}[g_{\tau \tau}j^\tau-g_{\tau x_3 }j^{x_3}]=0.
\end{equation}

The longitudinal conductivity reads
\begin{equation}
\label{eqav9}
\begin{split}
& \sigma_L (\omega,\overrightarrow{p},z)=\frac{j^{x_3}(\omega,\overrightarrow{p},z)}{F_{x_3 \tau}(\omega,\overrightarrow{p},z)},\\
& \partial_z \sigma_L=\frac{\partial_z j^{x_3}}{F_{x_3 \tau}}-\frac{ j^{x_3}}{F^2_{x_3 \tau}}\partial_z F_{x_3 \tau}.
 \end{split}
\end{equation}

The Kubo formula relates the longitudinal AC conductivity \(\sigma_L\) to the retarded Green's function,
\begin{equation}
\label{eqa12}
 \sigma_L(\omega)=-\frac{G^L_R (\omega)}{i \omega}.
\end{equation}

With the momentum limit \(P=(\omega,0,0,0)\), combining Eqs.~\eqref{eqav8} and \eqref{eqa91} gives the flow equation of the longitudinal channel
\begin{equation}
\label{eqav92}
 \partial_z \sigma_L=\frac{i\omega g_{xx_3}g_{zz}}{Q}(\sigma^2_L-\frac{Q^2(g^{\tau \tau}g^{xx_3}+g^{\tau x_3}g^{x_3 \tau})}{g_{xx_3}g_{zz}}).
\end{equation}

A similar procedure yields the flow equation for the transverse channel,
\begin{equation}
\label{eqav92}
 \partial_z \sigma_T=\frac{i\omega g_{xx_1}g_{zz}}{Q}(\sigma^2_T-\frac{Q^2 g^{zz}g^{\tau \tau}}{g^2_{xx_1}}).
\end{equation}

Imposing regularity at the horizon, we solve the equation subject to \(\partial_z \sigma_{L/T}(\omega,z)=0\), which yields the spectral function.
\begin{equation}
\label{eqa131}
 \rho(\omega)\equiv -Im G_R (\omega)=\omega Re \sigma(\omega,0).
\end{equation}

\begin{figure}[H]
    \centering
      \setlength{\abovecaptionskip}{0.1cm}
    \includegraphics[width=10cm]{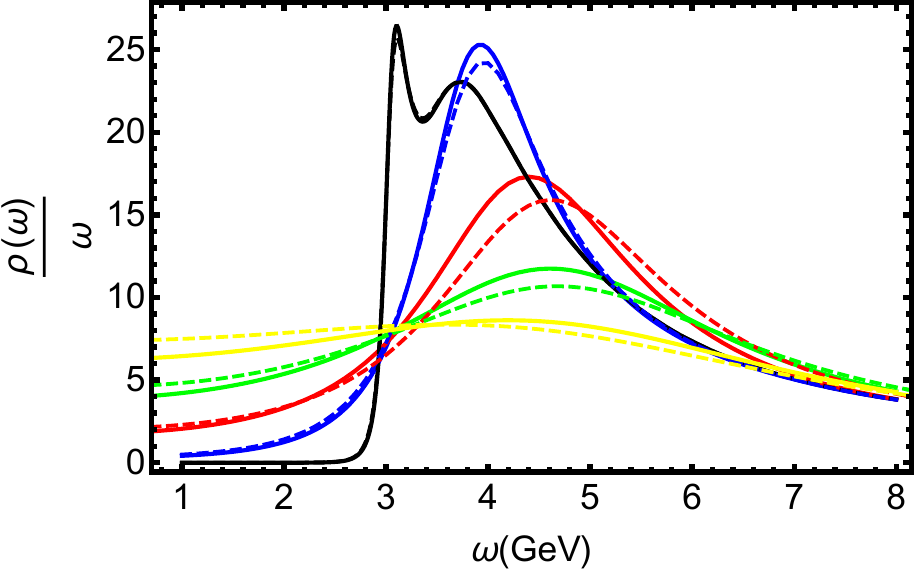}
    \caption{\label{fig1} Spectral functions for tetraquark $Z_c$ at $a=0.4$ for different values of temperature $T$. The black, blue, red, green and yellow line denote $T = 0.1,\ 0.4,\ 0.7,\ 1$, and $1.3$ GeV, respectively. The solid line (dashed line) represents the longitudinal (transverse) channel.}
\end{figure}

\begin{figure}[H]
    \centering
      \setlength{\abovecaptionskip}{0.1cm}
    \includegraphics[width=10cm]{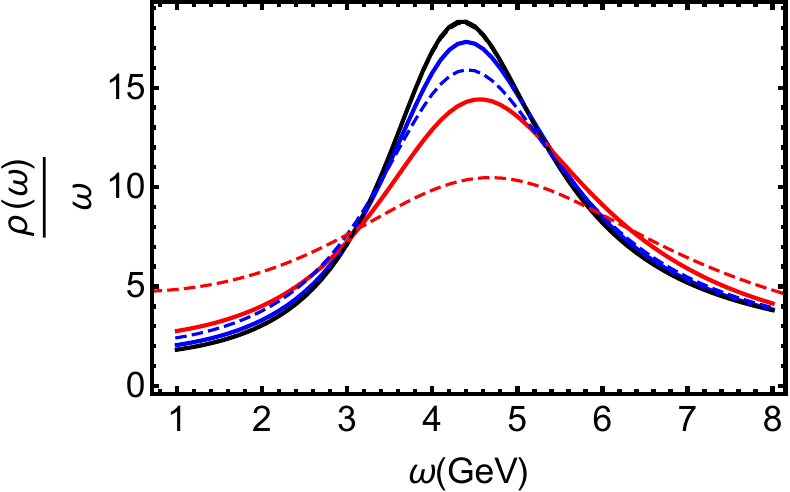}
    \caption{\label{fig2} Spectral functions for tetraquark $Z_c$ at $T=0.7$ GeV for different values of boost parameter $a$. The black, blue and red line denote $a = 0.1,\ 0.4$, and $0.7$, respectively. The solid line (dashed line) represents the longitudinal (transverse) channel.}
\end{figure}

To study how boost parameter affects the dissociation of exotic mesons, we compute the spectral functions in the spinning black hole background. For simplicity, we set \(L=1\). We work in the high-temperature regime, where the geometry reduces to a planar black brane \cite{Garbiso:2020puw}. Since the spinning black brane is stable for \(a < 0.75L\), we take boost parameter \(a=0.1, 0.4, 0.7\) in our calculations.

Fig.~\ref{fig1} shows the spectral functions of the tetraquark \(Z_c\) at boost parameter \(a=0.4\) for several temperatures. Each bell-shaped peak corresponds to a quasiparticle state. The first peak is the 1S state \(Z_c(3900)\), and the second is the 2S state \(Z_c(4200)\). As temperature \(T\) increases, both peaks broaden and decrease in height, which signals the progressive melting of the tetraquark. The heavier 2S state \(Z_c(4200)\) disappears near \(T=0.4\) GeV, while the 1S state \(Z_c(3900)\) survives until \(T=1.3\) GeV. The \(2S\) state melts earlier than the \(1S\) state. The dissociation is stronger in the transverse channel than in the longitudinal one, indicating that the boost parameter affects the two directions differently.

Fig.~\ref{fig2} presents the spectral functions of \(Z_c\) at \(T=0.7\) GeV for different values of the boost parameter \(a\). The width of a spectral peak is inversely related to the decay rate, and thus to the stability of the state. With increasing \(a\), the peaks become lower and broader in both the parallel and transverse channels. This behavior shows that boost parameter  enhances the dissociation of the tetraquark. The results of Figs.~\ref{fig1} and~\ref{fig2} indicate that boost parameter  promotes the melting of the tetraquark \(Z_c\) and the effect of boost parameter  is stronger in the transverse channel.

\begin{figure}[H]
    \centering
      \setlength{\abovecaptionskip}{0.1cm}
    \includegraphics[width=10cm]{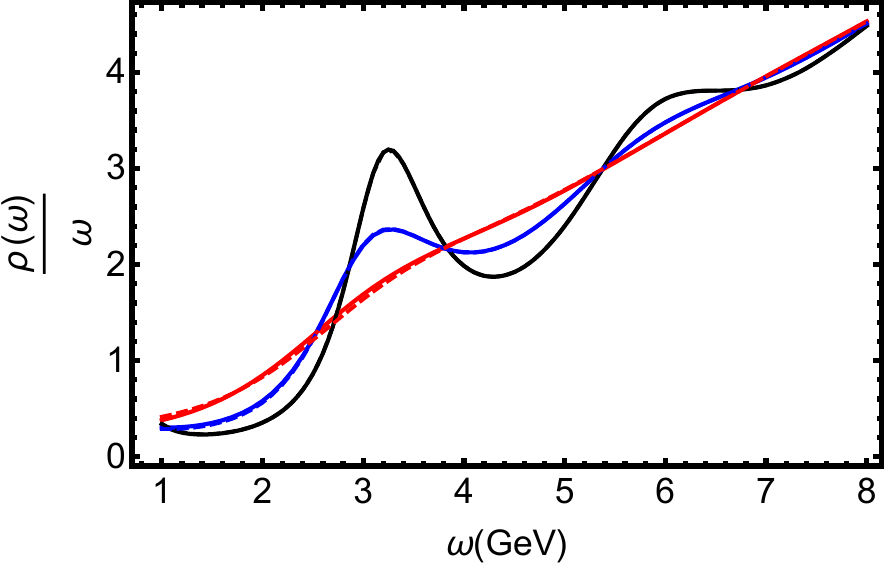}
    \caption{\label{fig3} Spectral functions for hybrid meson $\pi_1$ at $a=0.4$ for different values of temperature $T$. The black, blue, and red line denote $T = 0.1,\ 0.2$, and $0.3$ GeV, respectively. The solid line (dashed line) represents the longitudinal (transverse) channel.}
\end{figure}

\begin{figure}[H]
    \centering
      \setlength{\abovecaptionskip}{0.1cm}
    \includegraphics[width=10cm]{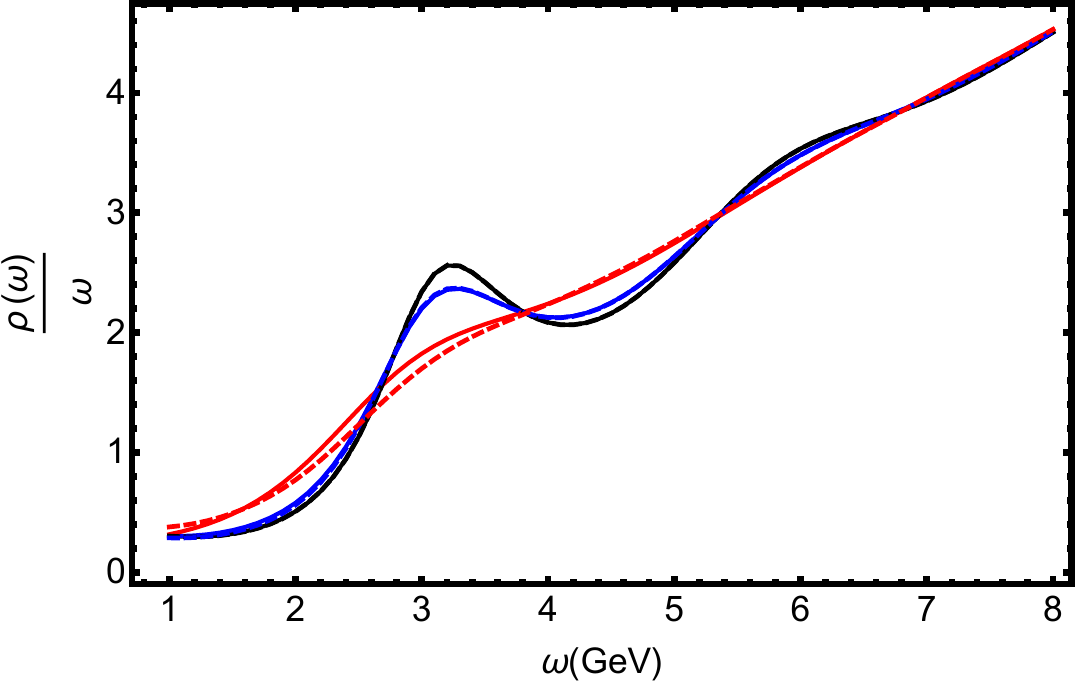}
    \caption{\label{fig4} Spectral functions for hybrid meson $\pi_1$ at $T=0.2$ GeV for different values of boost parameter $a$. The black, blue and red line denote $a = 0.1,\ 0.4$, and $0.7$, respectively. The solid line (dashed line) represents the longitudinal (transverse) channel.}
\end{figure}

Fig.~\ref{fig3} shows the spectral functions of the hybrid meson \(\pi_1\) at \(a=0.4\) for different temperatures. The first peak corresponds to the 1S state \(\pi_1(1400)\), and the second to the 2S state \(\pi_1(1600)\). As \(T\) increases, both peaks broaden and decrease in height, which signals the progressive melting of the hybrid meson. The heavier 2S state \(\pi_1(1600)\) disappears near \(T=0.2\) GeV, while the 1S state \(\pi_1(1400)\) survives until \(T=0.3\) GeV. The 2S state melts earlier. Dissociation is stronger in the transverse channel than in the longitudinal one, indicating that boost parameter affects the two directions differently. Fig.~\ref{fig4} presents the spectral functions of \(\pi_1\) at \(T=0.2\) GeV for different values of the boost parameter \(a\). With increasing \(a\), the peaks become lower and broader in both the parallel and transverse channels. This behavior shows that boost parameter  enhances the dissociation of the hybrid meson. The results indicate that boost parameter  promotes the melting of \(\pi_1\), and its effect is stronger in the transverse channel. Ref.~\cite{Zhu:2024uwu} investigated how boost parameter influences the dissociation of \(J/\psi\) and \(\Upsilon(1S)\) in the spinning black hole background. Our results show the same qualitative trend and are consistent with their findings.

We examine the thermal evolution of the \(\pi_1\) hybrid meson spectral function. From Fig.~\ref{fig3}, one finds that the \(\pi_1\) peak disappears completely at \(T=0.3\,\mathrm{GeV}\). In the planar limit, the background metric comes from the Myers--Perry solution. The authors of Ref.~\cite{Garbiso:2020puw} noted that the temperature is set by the horizon radius, with a large horizon corresponding to high temperature, and that the planar limit is the most reliable regime in large-horizon cases. They also emphasized, however, that most of their results are not restricted to the planar limit and were discussed for finite horizon radius. For our parameters, the horizon radius at \(T=0.3~\mathrm{GeV}\) is comparable to the AdS radius, still within the range considered there, and the boost parameter satisfies the stability condition. We therefore treat \(T=0.3~\mathrm{GeV}\) as the lowest temperature at which quantitative statements can be made safely, while \(T=0.1\) and \(0.2~\mathrm{GeV}\) serve only as qualitative references for the melting trend.

This is consistent with the recent review by Ref.~\cite{Amano:2023bhg}. In discussing quasinormal modes of rotating Myers-Perry black holes, they computed spectra for different values of the horizon and found that even at small to intermediate horizon radii, the lowest modes can still be computed reliably. They also noted that at lower temperatures the dispersion relations may deviate from integer power laws, yet fluid behavior persists with transport coefficients differing from boosted-fluid expectations. This is consistent with our use of \(T=0.3\,\mathrm{GeV}\) as the lowest temperature for quantitative statements. They further stressed that a rotating holographic fluid is not simply a boosted fluid and that the rotation effect is real \cite{Garbiso:2020puw}.

It is important to separate model-dependent numerical details from robust qualitative features. The precise value of the melting temperature, the detailed shape of the spectral peaks, and the dependence on the dilaton parameters are specific to the present soft-wall model and parameter choices. The qualitative picture, however, is stable within this model: the \(1S\) peak of \(\pi_1\) disappears earlier than that of \(Z_c\). Physically, \(\pi_1\) dissociating more easily than \(Z_c\) is consistent with the expectation that a gluon-excited state is less stable than a compact multiquark state. In the present model, this difference enters through the fitted Regge parameters of the two channels rather than through an explicit dynamical calculation of their binding energies. The model therefore captures the effective mass scales and spectral features of the two states, but it does not resolve their microscopic structure.

Finally, our main conclusions do not rely on the two low-temperature points. The monotonic melting from \(T=0.1\) to \(0.3~\mathrm{GeV}\) is continuous and supports the picture that \(\pi_1\) dissolves rapidly in a boosted quark-gluon plasma. The contrast that \(\pi_1\) fully dissolves within the reliable temperature range while \(Z_c\) survives to higher temperatures reveals the stability differences among exotic hadronic states in a spinning thermal medium, offering a useful reference for heavy-ion collisions.

\section{Conclusion and discussion}\label{sec:04}

We have investigated the dissociation of the \(Z_c\) and \(\pi_1\) in the spinning black hole background. Nonzero angular momentum is present in non-central heavy-ion collisions and may affect how exotic states behave in the medium. We compute the spectral functions in both the longitudinal and transverse channels, and study the roles of boost parameter and temperature in their melting.

From the results of spectral functions, one can find the boost parameter lowers the peak and widens it, thereby promoting the dissociation of \(Z_c\) and \(\pi_1\). The transverse channel responds more strongly than the longitudinal one. Temperature produces a similar effect and accelerates the melting. For both states, the 2S excitation dissolves before the 1S ground state. Within the present effective model, the fitted parameters for the \(\pi_1\) channel lead to a lower melting temperature than those for the \(Z_c\) channel. This ordering is consistent with the physical expectation that a gluon-excited state is less stable. The precise values of the melting temperatures depend on the dilaton parameters and should be read as model-dependent estimates; the qualitative ordering is the robust result. While the qualitative trends resemble those of conventional quarkonia, the quantitative response of the exotic channels is different, suggesting that boost parameter could serve as a probe to distinguish exotic states from ordinary quarkonia. In the planar limit the angular velocity is related to the boost parameter as \(\Omega = a/L^2\). Thus, the angular momentum promotes the dissociation of \(Z_c\) and \(\pi_1\).

These findings may offer insight into the dissociation of heavy and exotic mesons. Future work could include the configuration entropy in the spinning black hole background.

\section*{Acknowledgments}

Zhou-Run Zhu is supported by the National Natural Science Foundation of China under Grant No.12605236.

\section*{Data Availability Statement}

This manuscript has no associated data or the data will not be deposited.

\end{document}